\documentclass[a4paper]{article}

\usepackage{INTERSPEECH2020}

\title{StoRIR: Stochastic Room Impulse Response Generation \\ for Audio Data Augmentation}

\name{Piotr Masztalski, Mateusz Matuszewski, Karol Piaskowski, Michał Romaniuk}
\address{Samsung R\&D Institute Poland}
\email{\{p.masztalski, m.matuszews2, k.piaskowski, m.romaniuk2\}@samsung.com}

\begin{document}

\maketitle
\begin{abstract}
  In this paper we introduce StoRIR - a stochastic room impulse response generation method dedicated to audio data augmentation in machine learning applications. This technique, in contrary to geometrical methods like image-source or ray tracing, does not require prior definition of room geometry, absorption coefficients or microphone and source placement and is dependent solely on the acoustic parameters of the room. The method is intuitive, easy to implement and allows to generate RIRs of very complicated enclosures. We show that StoRIR, when used for audio data augmentation in a speech enhancement task, allows deep learning models to achieve better results on a wide range of metrics than when using the conventional image-source method, effectively improving many of them by more than 5 \%. We publish a Python implementation of StoRIR online \footnote{https://github.com/SRPOL-AUI/storir}.

\end{abstract}
\noindent\textbf{Index Terms}: room impulse response, data augmentation, speech enhancement, deep learning

\section{Introduction}

Deep learning tasks, like speech enhancement or acoustic event classification and localization, can significantly benefit from augmenting training datasets with room impulse responses (RIR) \cite{kinoshita2016reverb, Adavanne_2019, emmanou}. RIRs are a set of functions that describe the influence of a given acoustic environment when a sound wave propagates from a source to a receiver. Dataset augmentation with RIRs often leads to better generalization in real-life scenarios, where the acoustic environment has a large impact on the recorded audio signal. However, capturing real-life impulse responses requires a lot of time and resources and the openly available databases containing recorded RIRs can be insufficient for proper data augmentation. To address that issue, researchers have used computational methods that simulate RIRs with a lot of success \cite{Ko17-ASO, Hsi15-RSR}.  

\section{Background and related work}
\label{section:background}

\begin{figure*}[ht]
  \centering
  {\includegraphics[scale=0.9]{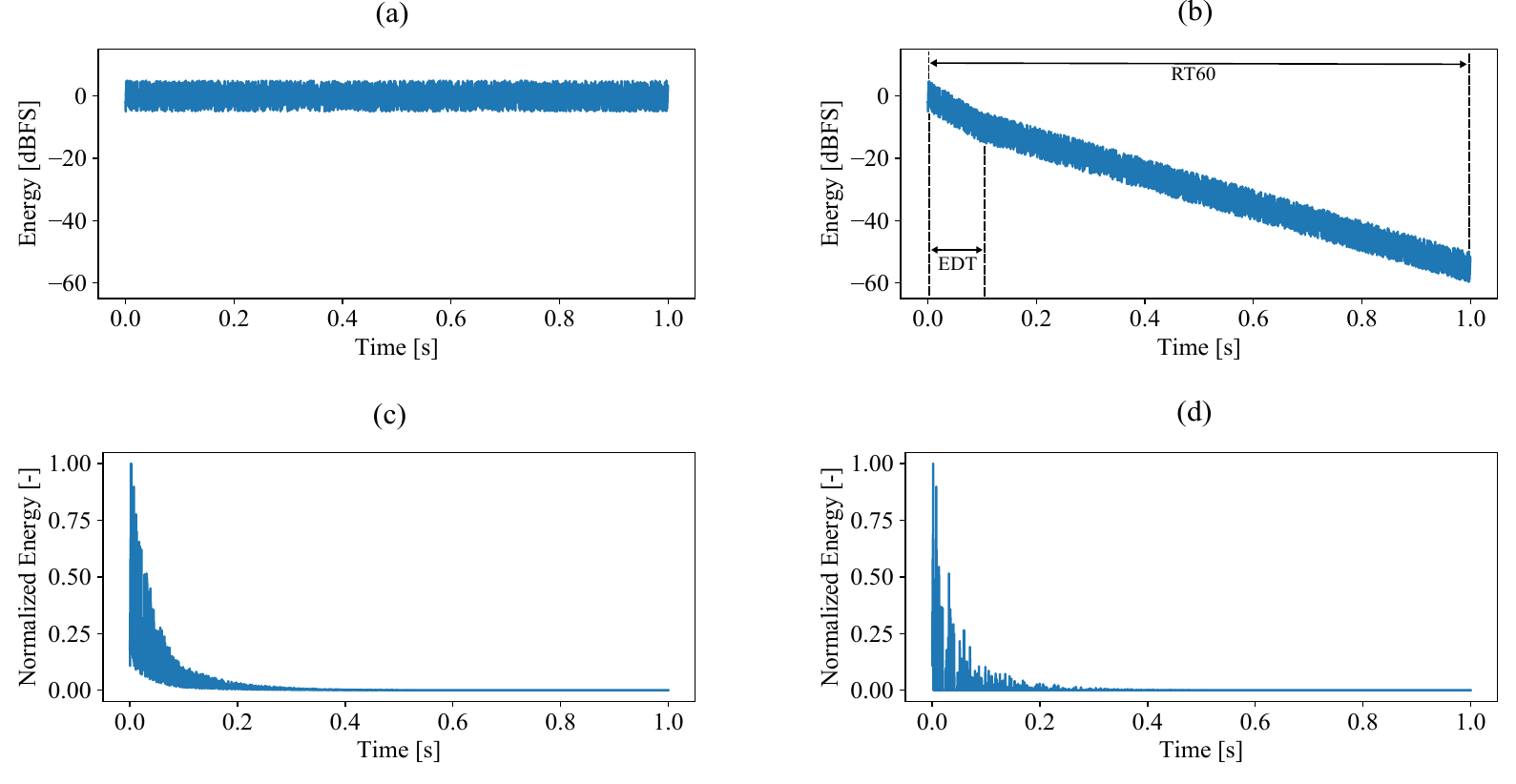}}
  \caption{Four steps of stochastic RIR generation. Noise generation - (a). Energy decay curve shaping - (b). Conversion from logarithmic to linear scale - (c). Adjusting energy distribution within the RIR - (d).}
  \label{fig:rir_generation}
\end{figure*}

There are several methods to calculate a RIR. The approach that is able to provide the most accurate results is based on numerically solving the wave equation (e.g. finite element method, boundary element method). However, these techniques are computationally very expensive and require a detailed mesh of the acoustic environment. Faster, but less accurate techniques are those based on the assumptions of geometrical acoustics (e.g. ray tracing , image-source) \cite{Savioja}. In geometrical acoustics, sound is assumed to propagate as rays, and all of its wave properties are neglected. This assumption is valid at high frequencies, where the wavelength of sound is short compared to surface dimensions and the overall dimensions of the space, but at low frequencies the approximation errors increase as wave phenomena play a larger role \cite{Savioja}. It is important to note, that for audio data augmentation in machine learning applications, the method of choice while simulating RIRs is almost exclusively image-source. We can attribute its popularity in this field to low computation complexity, that allows for online generation during training, and availability of open source software designed to model room acoustics using this method \cite{habets, Sch18-PRA}

Accuracy similar to that achieved by employing numerical methods in acoustics can be achievable in a much simpler way. Statistical Room Acoustics (SRA), under certain assumptions, can provide a statistical description of a RIR between a source and a receiver. Sabine \cite{sabine1923acoustics} presented the earliest attempt at room statistical methods. He introduced, in the late 19th century, a method for reverberation time calculation of a space without considering the details of its geometry. More than 50 years later, Schroeder \cite{schroeder1962frequency} extended Sabine’s fundamental work and derived a set of statistical properties describing the frequency spectrum of a random impulse response. Moorer et al. \cite{moorer1979reverberation} noted, that impulse responses in the finest concert halls around the world sounded remarkably similar to white noise with an exponential amplitude envelope. To test this observation, they generated synthetic impulse responses by shaping unit-variance Gaussian pseudo-random sequences with an exponential of the desired length. The direct sound was added by including an impulse at the beginning. Later, Polack \cite{polack1988transmission} developed a time-domain model excluding the contribution of the direct path and describing a RIR as a realization of a non-stationary stochastic process.

In this paper, we present a method of calculating RIRs based on SRA dedicated to machine learning data augmentation. To the knowledge of the authors this is the first application of SRA to machine learning tasks. The presented method is particularly useful for data augmentation, because it does not require defining room geometry and acoustic properties of walls, furniture, etc. which is necessary in other methods. Arguably, the fact that we do not define room geometry is a drawback for other purposes like auralization where the RIR of a specific room is needed. However, the purpose of data augmentation is to help machine learning models to generalize, so that they work in diverse acoustic environments. Therefore, using StoRIR, we generate RIRs based on several acoustic parameters that correspond to a class of rooms, without explicitly defining geometry or absorption coefficients.

\section{Proposed method}

\subsection{Room Acoustic Parameters}
\label{subsection:parameters}

There exist multiple parameters that describe acoustic properties of a room \cite{Kut20-RAF}. For the purpose of implementing StoRIR we use the following:

\begin{itemize}
\item Reverberation time (RT60) - a well-known, and probably the most common, parameter describing the acoustic behaviour of an acoustic cavity. RT60 quantifies the time it takes for the sound to decay by 60 dB after the sound source is removed. This term was introduced in Sabine’s pioneering research, in which he noticed that reverberation time was proportional to the volume of the room and inversely proportional to the amount of absorption. Because the absorptive properties of materials vary as a function of frequency, the reverberation time is also frequency dependent,
 \item Early Decay Time (EDT) - a very similar parameter to RT60. The difference relies in the amount of energy decay. EDT denotes how quickly sound energy drops by 10 dB, not 60 dB like in RT60. This is due to the fact that early reflections may decay with a different rate than the late part of the RIR, 
 \item Direct to Reverberant Ratio (DRR) - describes the energy ratio of the direct sound to the reverberant part of a signal,
 \item Initial Time Delay Gap (ITDG) - specifies the time between the arrival of the direct sound and the first reflection to the receiver.
\end{itemize}

\subsection{Stochastic impulse response generation}

The result of a geometric RIR generation method is an energetic impulse response (also called a reflectogram). We aim to simulate a similar outcome, but without providing the algorithm with room geometry, absorption coefficients and microphone/source placement. Hence, the method will not model a specific room but rather the impulse response itself. Normally, room acoustic parameters are obtained from a RIR. In our approach we reverse this property and generate the RIR from scratch by shaping it based on the acoustic parameter values. For the input to our algorithm we choose to consider RT60, EDT, DRR and ITDG described in detail in section \ref{subsection:parameters}. We argue, that using these paramaters can be more intuitive when generating a large amount of RIRs that should not necessarily correspond to a specific room geometry. This is often the case when augmenting data for machine learning purposes where we want models to generalize in a diverse range of acoustic environments.

The proposed stochastic RIR generation algorithm can be split into four steps: noise generation, energy decay curve shaping, conversion from logarithmic to linear scale and adjusting the energy distribution within the RIR (Figure \ref{fig:rir_generation}). 

\subsubsection{Noise generation}

We start constructing the RIR with generating an uniformly distributed noise vector $v$ of length $l$ equal to the RT60 parameter in samples. Both EDT and RT60 parameters describe the sound energy decay in decibels so, for convenience, we use the logarithmic scale in the beginning of RIR generation. Due to that, the range of the $v$'s uniform distribution will become the deviation range of reflection energies (in dB) in the resulting energetic impulse response. This range can be arbitrarily chosen or sampled from a distribution adding to the randomness of the generated RIR. We assume that the mean value of $v$ is 0 [dB].

\subsubsection{Energy decay curve shaping}

In order to obtain a simulated energy decay curve (${EDC}$), we introduce a negative slope to $v$ so that its mean value drops by 10 dB in time specified by the EDT parameter using

\begin{equation}
   \forall i \in 1, 2 ... k : EDC_i = v_i - \frac{10i}{k}
   \label{eq_edt}
\end{equation}

 where $k$ is the EDT parameter value in samples. \\
 
 After that, we concatenate the remainder of $v$ (decreased in level by 10 dB) with the $EDC$ vector and decrease it gradually by further 50 dB so that the overall level of $EDC$ drops by 60 dB in time specified by the RT60 parameter. We do it using the following equation:
 
 \begin{equation}
  \begin{split}
      \forall i \in k+1, k+2, ..., l : \\ EDC_i = v_i -\left[10 + \frac{50(i - k)}{l}\right]
   \label{eq_rt60}
  \end{split}
\end{equation}

\subsubsection{Conversion from logarithmic to linear scale}

The third step is converting $EDC$ to linear scale using

\begin{equation}
   EDC^{lin} = 10^{\frac{EDC}{10}}.
   \label{eq_decibels_to_gain}
\end{equation}

 After that, we normalize the resulting signal to its peak value. $EDC^{lin}$ can be interpreted as a maximally dense energetic impulse response, meaning energy is detected in every time slot (sample). We will refer to these energy peaks as reflections or rays.

\subsubsection{Adjusting energy distribution}

In order to account for the fact that, in real life, sound reflections reach the receiver with time gaps in between them, we make the obtained energetic impulse response $EDC^{lin}$ more sparse. We simulate delays between reflections by first eliminating a chunk of reflections of length equal to the ITDG parameter in samples after the initial sound ray. Then, we iteratively delete rays at random locations from the remaining part of $EDC^{lin}$ until we reach a desired DRR, hence the stochastic nature of the generated RIR. Analysis of sound propagation in enclosures shows, that in the early part of the RIR reflections are less densely distributed than in the reverberation tail \cite{Kut20-RAF}, therefore we assign a higher probability of deleting rays from the early reflection part of the RIR ($EDC^{lin}$) than from it's remainder.

A RIR generated in this manner does not represent any existing or modeled room but rather some imaginary room with desired acoustic parameters. Due to the randomness introduced in generating the energy decay curve from random noise, and the process of adjusting the energy ratios, the resulting RIR will differ with every generation even if we do not change the values of parameters it is based on. The more variability we allow in the acoustic parameters, the more the generated impulse responses will differ from each other. We can e.g. make a whole set of impulse responses where RT60 = 1 s, but with other parameter values sampled from chosen ranges, therefore creating an augmentation dataset of RIRs representing a class of virtual rooms with 1 second reverberation time.

\subsection{Other representations of the stochastic impulse response}

\begin{table*}[t]
  \caption{Results on Testset 1 (higher is better, bold text indicates best score per model within a given metric)}
  \label{tab:testset1_results}
  \centering
  \begin{tabular}{llllllll}
    \toprule
    \textbf{Model}  &  \textbf{Aug. method}&  \textbf{CSIG}&  \textbf{CBAK}&  \textbf{COVL}&  \textbf{fwSegSNR}  &  \textbf{STOI}  &  \textbf{PESQ} \\
    \midrule
    Noisy reverberant speech  &  - & 3.00 & 1.80 & 2.25 & 7.1 & 0.62 & 1.91         \\
    \midrule
    \midrule
    \multirow{3}{*}{DenoiseNet} &  No augmentation& 2.87 & 1.87 & 2.20 & 7.1  &  0.63 &  2.09  \\
                               &  Image-Source & 3.63 & 2.07 & 2.73 & 9.2 & 0.67 & 2.40  \\
                               & StoRIR & \textbf{3.66} & \textbf{2.08} & \textbf{2.75} & \textbf{9.3}   &  \textbf{0.69} & \textbf{2.41}  \\
    \midrule
    \multirow{3}{*}{DereverbNet} &  No augmentation & 3.69 & 2.26 & 2.98 & 10.2 &  0.64 &  2.07  \\
                                &  Image-Source & \textbf{3.97} & \textbf{2.45} & \textbf{3.26} & \textbf{11.3} & \textbf{0.76} & 2.34 \\
                                &  StoRIR & 3.95 & 2.35 & 3.20 & 11.2 & 0.70 & \textbf{2.45}         \\
    \bottomrule
  \end{tabular}
\end{table*}

\begin{table*}[t]
  \caption{Results on Testset 2 (higher is better, bold text indicates best score per model within a given metric)}
  \label{tab:testset2_results}
  \centering
  \begin{tabular}{llllllll}
    \toprule
    \textbf{Model}  &  \textbf{Aug. method}&  \textbf{CSIG}&  \textbf{CBAK}&  \textbf{COVL}&  \textbf{fwSegSNR}  &  \textbf{STOI}  &  \textbf{PESQ} \\
    \midrule
    Noisy reverberant speech  &  - & 3.25 & 1.87 & 2.44 & 8.1  & 0.77 & 2.01          \\
    \midrule
    \midrule
    \multirow{3}{*}{DenoiseNet}    &  No augmentation& 3.17 & 1.96 & 2.42 & 7.9  &  0.74 &  2.08  \\
                                   &  Image-Source& 3.87 & 2.08 & 2.89 & 9.9  &  0.83 &  2.53  \\
                                   &  StoRIR & \textbf{4.02} & \textbf{2.23} & \textbf{3.05} & \textbf{10.6}   &  \textbf{0.84} &  \textbf{2.59}  \\
    \midrule
    \multirow{3}{*}{DereverbNet}  &  No augmentation& 3.69 & 2.29 & 2.98 & 10.0 &  0.77 &  2.03  \\
                                  &  Image-Source & 4.20 & 2.46 & 3.40 & 11.7 & 0.81 & 2.55  \\
                                  &  StoRIR & \textbf{4.34} & \textbf{2.51} & \textbf{3.60} & \textbf{12.6}     & \textbf{0.86} & \textbf{2.62} \\
    \bottomrule
  \end{tabular}
\end{table*}

After executing all of the steps described above, we obtain an energetic impulse response that can be directly convolved with an arbitrary audio signal resulting in a reverberation effect. However, the reverberation time of the space represented by this RIR is equal across the whole frequency spectrum which, most of the time, is not the case with real impulse responses due to stronger air and surface attenuation of sound in higher frequencies \cite{Kut20-RAF}. To account for that, the RIR can be generated separately for each 1/1 or 1/3 octave band with decreasing reverberation time. Another modification can be to generate the RIR with lower time resolution (e.g. 5 ms) and then adapt it to the sampling frequency using the Poisson-distributed noise method described in \cite{Sch11-PBR} (however, it should be noted that this method requires an estimation of the room volume). In our experiments we decided to use the most basic version of the energetic impulse response without any of the above described modifications, as we have experimentally established that it produces the best results in our evaluation task.

\section{Experimental evaluation}

\subsection{Model description}

Although our method can be employed in virtually any task that can benefit from RIR augmentation, we decided to evaluate it on a speech enhancement task. A large amount of research has been conducted in the field of noise suppression in anechoic conditions \cite{pascual2017segan, Choi18-PAS, macartney2018improved, park2016fully}. However, these experimental scenarios are not applicable to real-life use cases, as the acoustic characteristics of the space where the sound sources are placed cause a significant difference to the spectral structure of resulting audio signals. In order to account for this difference, it is common to convolve noisy audio with various RIRs so that the model can suppress noise and reverberation at once. To evaluate the proposed augmentation method we use one deep learning model designed for denoising \cite{Choi18-PAS}, and one designed for dereverberation \cite{Wang20-DLB} which we call DenoiseNet and DereverbNet. Both are based on the U-Net \cite{Ron15-UNC} architecture with DenoiseNet using complex-valued operations, and DereverbNet employing DenseNet-like blocks \cite{Hua17-DCC} and two LSTM layers in the bottleneck. Both models are used for the exact same task of simultaneous speech denoising and dereverberation.

\subsection{Dataset}

\subsubsection{Training}

For all experiments, we use the clean and noisy parallel speech database by Valentini et al. \cite{Val16-SEF} which consists of around 400 utterances from each of 14 male and 14 female speakers. The authors obtained the training dataset by mixing speech files from the Voice Bank corpus \cite{Vea13-TVB} with 10 different types of noise (2 artificial, and 8 from the DEMAND database \cite{Thi13-TDE}) at four different SNRs (15 dB, 10 dB, 5 dB and 0 dB). To evaluate the performance of our RIR generation method, we compare it with the image-source method. We generate 50,000 impulse responses with the image method using Pyroomacoustics \cite{Sch18-PRA}, an open source Python package for room acoustics simulation. The details of simulated room geometries and placement of sound sources and microphones are the same as described in \cite{Wang20-DLB}. The one difference is, we change the desired reverberation time range to 0.2 s - 0.7 s (it has to be converted to absorption coefficients using the Sabine's formula for the purpose of using the image-source method). For comparison, we also generate 50,000 equivalent RIRs with the proposed method setting the RT60 range to 0.2 s - 0.7 s, the EDT range to 50 ms - 100 ms, the ITDG range to 3 ms - 10 ms and the DRR range to -7 dB - 0 dB. During training the generated RIRs are randomly convolved with noisy speech utterances to obtain noisy and reverberant signals. It is important to note, that during training, only simulated RIRs were used without adding any real-life recorded ones.

\subsubsection{Testing}

The performance of each model is evaluated on two test sets. The first one (Testset 1) is an unchanged version of the noisy and reverberant test set proposed by Valentini \cite{Val17-NRS}. Because of the fact that this dataset covers only 3 room configurations (two of which are recorded in a somewhat artificial setting using a room with configurable reverberation time), we create a second test set (Testset 2) utilizing the same underlying clean and noisy files, but convolved randomly with 12 RIRs from 6 different rooms (two offices, two meeting rooms, a lecture room and a building lobby) in order to further evaluate generalization in real-life settings and more complex room geometries. The RIRs used in Testset 2 are obtained from the ACE challenge corpus \cite{Eat15-TAC} and their reverberation time ranges from 0.3 s to 0.75 s. All RIRs in both test sets are real-life recordings.

\subsection{Data preprocessing}

The original audio signals were first downsampled from 48kHz to 16 kHz (which is a common practice in speech enhancement tasks) and then, for the final model input, complex-valued spectrograms were obtained from raw waveforms. The STFT was computed with a 64 ms window and a 16 ms hop size for DenoiseNet, and a 32 ms window and 8 ms hop size for DereverbNet as stated in the original papers. The training target for all experiments was clean anechoic speech.

\subsection{Evaluation metrics}

To evaluate the performance of the speech enhancement models, and hence the viability of data augmentation using the proposed method, we choose six objective speech quality and intelligibility metrics: CSIG - mean opinion score (MOS) predictor of signal distortion, CBAK - MOS predictor of background-noise intrusiveness, COVL - MOS predictor of overall signal quality, fwSegSNR - frequency-weighted segmental signal to noise ratio, STOI - short time objective intelligibility and PESQ - perceptual evaluation of speech quality. For calculating metrics we used code repositories available online \footnote{https://github.com/IMLHF/Speech-Enhancement-Measures} \footnote{https://github.com/mpariente/pystoi}
\footnote{https://github.com/vBaiCai/python-pesq}.

\section{Results and discussion}

Tables \ref{tab:testset1_results} and \ref{tab:testset2_results} summarize the metric scores on both test sets. It is worth to mention the substantial boost of performance caused by both RIR augmentation techniques when comparing with no augmentation. In some cases, models trained with no augmentation achieve even worse scores than the original noisy and reverberant mixture. Both RIR generation methods lead to improved speech quality metrics, however, when comparing the two, using StoRIR produces better results in more cases overall. The scores are improved on Testset 2 across all metrics, which we see resulting from better representation of unconventional room geometries (like building lobby) by our RIR generation method. The 0.9 dB improvement in fwSegSNR on DereverbNet seems to be the most impressive result on this testset. More comparable results on Testset 1 may result from the fact that two out of three RIRs used in this test set are recorded in a configurable reverberation room with a perfect ShoeBox geometry, identical to the ones modeled with the image-source method. Still, even though the image-source method has a somewhat unfair advantage in this test scenario, StoRIR manages to provide similar or better results (especially on DereverbNet where the PESQ improvement reached 0.11). The presented results also show that the performance of our method is model agnostic.
\section{Conclusions}

In this paper we proposed StoRIR, a method for room impulse response generation that does not require any information about room geometry, absorption coefficients or microphone and sound source placement. We show that when used for data augmentation, our method substantially improves the performance of deep learning models employed for a speech enhancement task in reverberant conditions. When compared with the image-source method for RIR generation, StoRIR achieves superior results when dealing with real reverberation in a wide range of acoustic environments.



\bibliographystyle{IEEEtran}

\bibliography{mybib}

\end{document}